\documentstyle[11pt,paspconf,epsf]{article}

\begin{document}

\title{The Redshift Cutoff and the Quasar Epoch}

\author{P. A. Shaver,  I.M. Hook}
\affil{European Southern Observatory, Karl-Schwarzschild-Str. 2,
D-85748 Garching bei M\"unchen, Germany}

\author{C. A. Jackson}
\affil{Institute of Astronomy, Madingley Road, Cambridge CB3 OHA, UK;
School of Physics, University of Sydney, Sydney, NSW 2006, Australia}

\author{J. V. Wall}
\affil{Royal Greenwich Observatory, Madingley Road, Cambridge CB3 OEZ,
UK}

\author{K. I. Kellermann}
\affil{National Radio Astronomy Observatory, Edgemont Road,
Charlottesville, VA 22903-2475, USA}

\begin{abstract}
We have obtained complete redshift information for a sample of 442
radio-loud quasars with flux densities $S_{11} \geq 0.25$
Jy. These come from a completely-identified sample of 878
flat-spectrum radio sources, so there is no optical magnitude limit.
With these quasars, therefore, we can map out the entire quasar epoch,
unhindered by optical selection effects and any intervening dust.
The rapid decline in true space density above $z \sim 2.5$ 
is confirmed. It agrees closely with that obtained from
optically-selected samples, indicating that these are also not significantly
affected by dust. The evolution of the quasar space density is similar
to that deduced for the global star formation rate, implying that 
the quasar epoch may also be that of star formation. This sample of 
radio-selected quasars can now be used in a
variety of other studies, including unbiased searches for damped Ly$\alpha$
absorbers, studies of the phenomenon of ``red quasars'', and searches 
for high-redshift molecular absorption systems.
\end{abstract}

% (Keywords have to be included, but are not printed in the hardcopy.)

\section{Introduction}

Radio samples offer one of the best possibilities at the moment for
searching for high redshift quasars and obtaining complete samples of 
quasars that are free of optical selection effects. Radio surveys 
cover vast areas of sky, providing thousands of candidates with accurate 
positions. If these are flat-spectrum sources, the probability that 
they are quasars is very high and the selection effects relatively 
simple and well understood. In particular, there is no bias against high
redshift quasars, at least out to $z \sim 10$. \\

In 1992 we initiated a study aimed at the complete identification of 
a sample of flat-spectrum radio sources from the Parkes catalogue. The sample
was comprised of all sources from the catalogue with spectral indices
between 11 and 6 cm flatter than -0.4, in the declination range
+2.5$^{\circ}$ to -80$^{\circ}$ (excluding low galactic latitudes ($|b|
< 10^{\circ}$) and regions around the Magellanic Clouds. Of these 878
sources, only 229 were listed in the catalogue as unidentified. For
these we obtained accurate radio positions using new VLA and
Australia Telescope measurements and previously published and
unpublished measurements. These sources were then identified, either
from cross-correlation with the COSMOS catalogue of the southern sky
(based on the UK Schmidt J survey with a limiting magnitude of
$\sim$ 22.0-22.5), or from CCD observations made with the ESO 3.6m
telescope on La Silla. We now have a completely identified sample of
878 sources, with no optical magnitude limit or selection effects.
In addition, we have obtained complete spectroscopic data on a
sub-sample of 442 stellar identifications - quasars and BL Lacs. \\

This sample can be used for a variety of purposes. Our primary
objective was to search for $z > 5$ quasars, and test the reality of
the apparent redshift cutoff (Shaver {\it et al.}, 1996a). 
It also becomes possible to map out the
entire ``quasar epoch'' with one single sample, free of the
complications of optical selection effects, to study the radio and
optical luminosity functions as a function of redshift, and to set
limits on the obscuration of the distant Universe by dust. It can be
used for unbiased searches for damped Ly$\alpha$ systems, studies of
``red quasars'', and candidates for molecular absorption systems. 
We briefly describe some of these aspects below. Full accounts of this
work will be published in a series of three papers (Shaver {\it et
al.,} 1998; Jackson {\it et al.,} 1998; Hook {\it et al.,} 1998).

\section{``The Edge of the World''}

It was already becoming apparent in the late 1960s that the rapid
increase in the number of quasars with redshift did not continue
beyond $z \sim 2$. indeed, several authors suggested that there
could be a redshift cutoff -- a real decrease in the space density of
quasars which was not merely due to observational selection effects. 
Sandage (1972) wrote {\it ``The possibility that we have already seen 
to the edge of the world is quite remarkable. To prove that we have 
would be quite unique.''} Over the ensuing years, however, the reality of a 
redshift cutoff has been difficult to establish with certainty. \\

Given their great luminosities, quasars could easily be seen out to
very high redshifts ($>> 10$), if they are there. Any search should
ideally be open to the highest possible redshifts, and not limited
by the search technique. A radio-based search is the obvious choice at
present. Not only is it free of the selection effects which limit
optical searches, it is also free of any obscuration by dust. \\

We therefore used our large complete sample to search for high
redshift quasars. Quasars at very high redshifts ($z > 5$) can be 
distinguished by the
fact that they should be heavily obscured in the optical $B$ band by
intervening hydrogen, because the Lyman limit is redshifted past this
band, and they will only be detectable at very red wavelengths. Thus,
our strategy was to look for extremely red stellar objects at the
positions of our flat-spectrum radio sources. \\

There were two critical assumptions. (1) Very high-redshift quasars 
will still have flat spectra between the observed wavelengths 11 and 
6 cm. However, spectra which are flat at the usual centimeter wavelengths 
often become steeper at shorter wavelengths, and at sufficiently high 
redshifts this steep portion of the spectrum will be redshifted into 
the window between 11 and 6 cm where the sample is selected, so that 
very high-redshift quasars could be excluded in the selection process. 
Examination of the quasi-simultaneous radio spectra of a sample of quasars
and BL Lac objects measured by Gear {\it et al.,} (1994) indicates
that this effect only becomes important above $z = 10$. (2) $z > 5$ 
quasars will be obscured in the optical $B$ band. This is justified 
on the basis of observed quasar spectra at redshifts up to 5 
(Storrie-Lombardi {\it et al.,} 1996) and extrapolations of the 
statistics of intervening hydrogen absorbers to higher redshifts 
(M{\o}ller (private communication); M{\o}ller \& Warren, 1991). 
10 magnitudes of extinction is typical 
in the $B$-band, even at $z = 5$. \\

All of the sources in this sample have been optically identified, 
either as galaxies or as stellar objects which are present in the
$B$-band. In neither case can they be quasars at $z > 5$. The only
source that came close was PKS 1251-407, identified with a very red 
stellar object, although because it was still faintly visible in $B$ 
it was not considered a $z > 5$ candidate. Indeed, it was found to have 
a redshift of 4.46, making it the highest-redshift radio-loud quasar 
known at the time (Shaver {\it et al.}, 1996b). Objects at higher 
redshifts could have been found with similar ease. Thus, as all 
sources are identified, there are simply none left which could be quasars 
at $z > 5$. \\

This enables us to place an upper limit on the space density of flat-spectrum
radio-loud quasars at $z > 5$ {\it which is independent of optical magnitude}. 
We consider only those areas which were completely surveyed down to 
0.25 Jy at 11 cm. The absence of any quasars in the comoving 
volume covered by this 3.8 steradian area and the redshift range $5 < z 
< 7$ can be directly compared with measured space densities in lower 
redshift bins of quasars with radio luminosities $P_{lim} \geq 1.1$ 10$^{27}$ 
W Hz$^{-1}$ sr$^{-1}$ corresponding to the flux limit at 
$z = 7$ ($H_{\circ}$ = 50 km s$^{-1}$ Mpc$^{-1}$ and $q_{\circ}$ = 0.5
are used throughout this paper). As we can use only sources with known
redshifts in the lower redshift bins, some ({\it eg.} BL Lacs) are
excluded,  so the space densities we compute at $z < 4$ are {\it
lower} limits, to be compared with the {\it upper} limit at $z > 5$. \\

This upper limit, normalized to the space densities found at $z \sim 2-3$,  
is marked at $z = 6$ in fig. 1. The pronounced decrease at high redshift is
obvious; considering that 25 quasars above $P_{lim}$ are found 
at $z < 4$ and none at $5 < z < 7$, the turnover is significant at the
99.96\% level (99.98\% when we use our entire sample, including zones 
of the Parkes catalogue complete to 0.20 and 0.65 Jy). If there were no
fall-off in the space density of quasars at high redshift, we would
expect to find at least 15 quasars at $5 < z < 7$ in our sample. \\

This absence of high-redshift quasars in our sample is very unlikely 
to be due to obscuration:

\begin{enumerate}
 \itemsep=-3pt

 \item Radio emission is unaffected by dust, 
and all sources are already accounted for - there are simply none left
to be $z > 5$ quasars. This excludes the possibility of obscuration by
intervening galaxies or dusty Lyman limit systems too faint to be seen
directly. 

 \item The possibility that some of the sources 
are in fact at $z > 5$ but have been {\it misidentified} with 
visible foreground galaxies that obscure them cannot explain the turnover, 
because it would require accurate positional coincidences of {\it all} 
of the $z > 5$ quasars with observed foreground galaxies. 
The probability of that is negligible, about 10$^{-15}$. 
Furthermore, gravitational bending of the light from the 
quasar by the intervening galaxy would generally increase the impact 
parameter and decrease any obscuration (Bartelmann \& Loeb, 1996). 

 \item Another 
possible source of obscuration at high redshifts is Thomson scattering
by an ionized intergalactic medium, which affects all frequencies 
$\ll mc^{2}/h$ equally. However, even if all the baryons in the Universe 
were distributed in an ionized intergalactic plasma with $\Omega_{B}$
at the improbably high value of 0.5, the effect on the results shown 
in fig. 1 would still be small. 

 \item The possibility that some sources have
been {\it completely} missed (also in the radio) due to both optical
obscuration and radio free-free absorption in intervening objects is highly
implausible, because a large free-free optical depth would imply a
small size for any reasonable excitation parameter, hence a very small
probability of a line-of-sight coincidence. 

 \item Obscuration by large
amounts of dust and/or ionized gas {\it intrinsic} to the quasar 
is conceivable - it could result in only a faint red galaxy being
observed (although at $z > 5$ it would likely be below our detection 
limit), or nothing at all (even in the radio). However, this
possibility does not affect the conclusions drawn here. We are
concerned with a particular observed phenomenon, {\it i.e.} quasars
similar to those seen at $z \sim 1-2$, and the space density of such objects
decreases at high redshifts. There must obviously be precursors at
still higher redshifts, at least in the form of mass concentrations,
but that is a different issue. Furthermore, even if there are
``hidden'' quasars enshrouded in dust and/or ionized gas, if quasars are
short-lived objects we would expect such hidden quasars to exist at {\it
all} redshifts, so the redshift turnover would be unaffected ({\it
eg.} Webster {\it et al.} (1997) find that the incidence of ``red 
quasars'' is not redshift-dependent). 

\end{enumerate}

For all of these reasons, it is concluded that the observed turnover
in the space density of flat-spectrum radio-loud quasars is real, and
not due to obscuration. The actual number of high redshift quasars 
must be very low: we expect that the number of flat-spectrum quasars 
at $z > 5$, even down to $S_{6} = 50$ mJy, is less than 30 over the whole sky.

\begin{figure}[h]
%\begin{figure}[p]
\plotfiddle{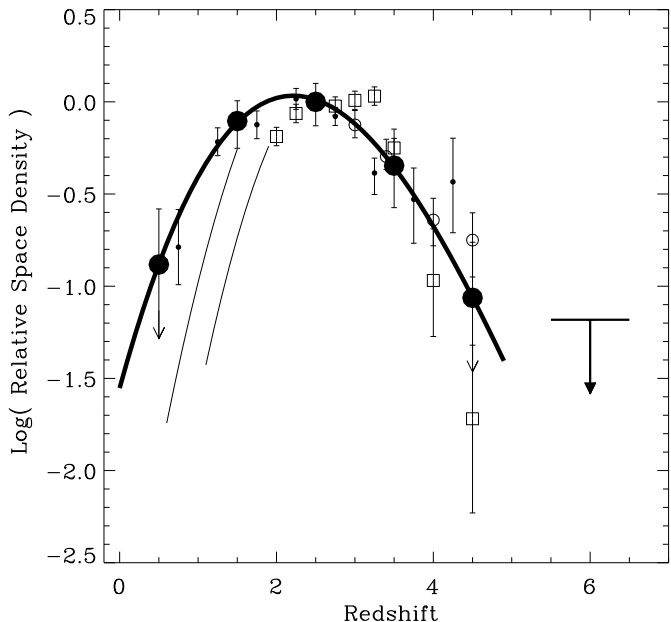}{7.0cm}{0.0}{75.0}{75.0}{-180.0}{-280.0}
%\plotone{dust.ps}
\caption{Space densities, normalized to $z \sim 2-3$ and plotted as a 
function of redshift, for the Parkes flat-spectrum radio-loud quasars 
with $P_{11} > 7.2$ 10$^{26}$ W Hz$^{-1}$ sr$^{-1}$ ({\LARGE $\bullet$}). 
The thick solid line is an approximate fit to these data. The upper 
limit shown at $z = 6$ pertains to the results 
given in section 2 of this paper. For comparison, similarly normalized
space densities are also shown for the optically-selected quasar samples 
of Warren {\it et al.} (1994) ($\Box$), Schmidt {\it et al.} (1995) 
({$\circ$}), and Hawkins \& V\'{e}ron (1996) ($\bullet$). The thin 
solid lines represent luminosity functions used by Warren {\it et al.}
and Schmidt {\it et al.} as lower-redshift continuations of their 
high-redshift space densities.}
\label{dust}
\end{figure}

\section{The Quasar Epoch}

As we now have complete identifications for our radio sources, once we
have redshifts we can map out the entire ``quasar epoch'' using just 
one complete sample with no optical magnitude limit, selection effects, 
or complications due to dust obscuration. We selected a sub-sample of 
442 stellar identifications (quasars and BL Lacs) with 
$S_{11cm} \geq 0.25$ Jy in the range $+2.5^{\circ} > \delta > 
- -45^{\circ}$. Spectroscopic information was already available for 
two-thirds of these sources, and it was only necessary to obtain spectra 
of the others. This has recently been completed, using the ESO 
3.6m telescope. \\

Of these 442 sources, 91\% are quasars and the rest Bl Lacs. 
Detailed luminosity functions will be published in a later paper; here
we concentrate on the redshift distribution of the most luminous 
quasars, those with $P_{11} \geq 7.2$ 10$^{26}$ W Hz$^{-1}$ sr$^{-1}$, 
corresponding to $\geq 0.25$ Jy at $z = 5$. These are indicated 
by the large filled circles in fig. 1, and confirm the conclusion drawn
above that there is a real decrease in the space density at high
redshifts. \\

A striking feature of fig. 1 is that the 
optically-selected samples show the same turnover as these luminous
radio-selected quasars. As the radio sample
is unaffected by dust obscuration, this strongly suggests that the
turnover in the optical samples is also due to a decline in the 
true space density rather than to dust obscuration as had been 
proposed earlier (Ostriker \& Heisler, 1984). Other arguments for 
the universality of the redshift turnover can be made:

\begin{enumerate}
 \itemsep=6pt

 \item We implicitly assume that the true ratio of radio-loud to 
 radio-quiet quasars remains constant with redshift. This is 
 the simplest assumption, which should therefore be preferred 
 in the absence of any compelling counter-arguments, theoretical or
 observational. The only relevant observational evidence supports this
 hypothesis: the fraction of optically-selected quasars which are 
 flat-spectrum radio sources appears to be roughly constant with 
 redshift at $z > 1$ (Hooper {\it et al.}, 1995)

 \item If the turnover in optically-selected samples were due to
obscuration, radio quasars would be similarly affected. Thus, if the 
apparent space density were down by a factor 
of, say, ten at $z \sim 4.5$ due to obscuration, there would be only a 10\%
chance of our having found PKS 1251-407 rather than a blank field. 

 \item The ultraviolet background radiation field may have a similar
dependence on redshift, as would be expected if it is produced 
largely by the quasars; it increases by two orders of magnitude from
$z \sim 0.5$ to $z \sim 3$, and may decrease again at still higher
redshifts (Bechtold (1995) and references therein), although this
remains uncertain.

 \item Recent studies (Fall \& Pei, 1993; Pei \& Fall, 1995) which 
consider the possible effects of dust in 
intervening galaxies and HI clouds suggest that the obscuration due to
these objects would not suffice to produce the redshift turnover. The
exact amount of dust in such objects is a matter of
debate ({\it eg.} Prochaska \& Wolfe, 1996; Lu {\it et al.}, 1997; 
Pettini {\it el.}, 1997), but in any case it is not thought to be very large.

\end{enumerate}

The results presented in fig. 1 set limits 
on the {\it total} intervening dust obscuration along the line of sight 
- - intergalactic as well as that from intervening galaxies and damped 
Ly$\alpha$ absorbers. Again making the simplest assumption, namely 
that the radio fraction of quasars remains constant at the highest 
redshifts, the close agreement in the space density turnover for the 
radio and optical samples indicates that the effect of obscuration by
intervening dust is small. \\

\begin{figure}[ht]
\plotfiddle{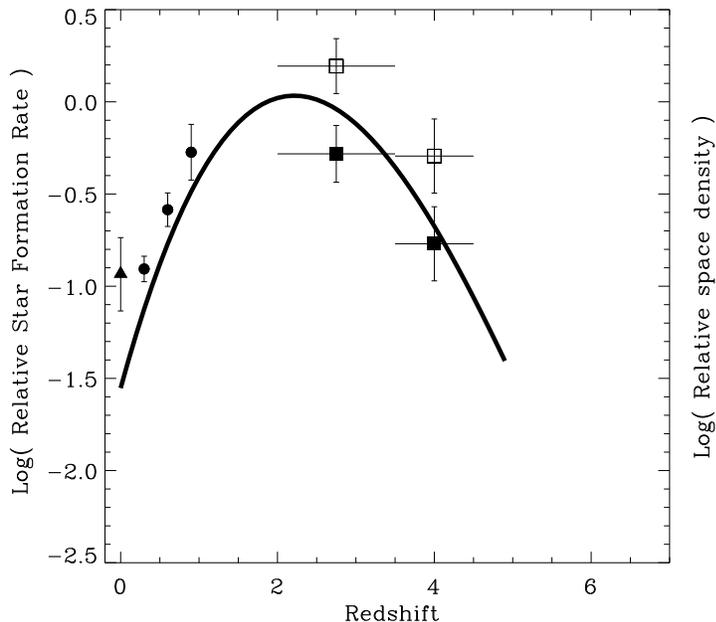}{7.0cm}{0.0}{75.0}{75.0}{-180.0}{-280.0}
%\plotone{sf.ps}
\caption{Evolution of the star formation rate (reproduced from Pettini
{\it et al.}, 1997), compared to that of the space density of quasars 
(solid line, from fig. 1). The filled squares are measurements from
the HDF, the circles from the CFRS and the triangles from a local
H$\alpha$ survey. The open squares show the HDF values corrected for
dust extinction in the galaxies.}
\label{sf}
\end{figure}

Is the quasar epoch coextensive with the epoch of star formation?
Fig. 2 shows the evolution of the space density of quasars compared
with that of the star formation rate as deduced from recent deep ground-
and space-based observations of faint galaxies.
They are remarkably similar. However, the star formation rates have 
to be corrected for the possible effects of dust within the galaxies.
The corrections can be severe especially at high redshifts, because
the relevant observations are made in the rest-frame ultraviolet. Thus
the redshift of the peak, and the decline to higher redshifts, are not
yet well determined. Infrared and submillimeter observations will
ultimately be required to settle these issues. Nevertheless, first
indications are that the quasar epoch may indeed also be that of star
formation. \\

\section{Red Quasars}

A small fraction of the sources in our sample are so-called ``red
quasars''. These objects, discovered by Rieke {\it et al.} (1979), 
have an extremely red nucleus - either an infrared excess due to 
synchrotron emission or an optical deficiency due to instrinsic dust
absorption. They may appear as faint galaxies in the optical, and as 
quasars in the near-infrared. The fraction of such objects in our
total sample of 878 sources is probably less than about 10\%. As noted above, 
these do not affect our conclusions concerning the redshift turnover. \\

Selecting the reddest of our stellar identifications, we find that
about one-quarter are BL Lacs, compared to 9\% with no colour selection, 
supporting the synchrotron interpretation for many of the red objects, as
originally proposed by Rieke {\it et al.} (1979). However, the discovery 
of molecular line emission from a sample of red quasars by Carilli 
{\it et al.} (1998) indicates the likely presence of dust in some 
cases, possibly concentrated in dense nuclear tori. Our sample will be
used to study the phenomenon of red quasars, and to search for more candidates
for this new and exciting area of molecular astrophysics.

\end{document}